\documentclass[preprint,showpacs,secnumarabic,amssymb,floatfix,nobibnotes, aps, prb]{revtex4}

\usepackage{amsmath}
\usepackage{graphicx}
\usepackage{epstopdf}

\newcommand{\bra}[1]{\langle #1|}
\newcommand{\ket}[1]{|#1\rangle}
\newcommand{\scalar}[2]{\langle #1|#2\rangle}
\newcommand{\Equation}[1]{Eq.~(\ref{#1})}
\newcommand{\Equations}[2]{Eqs.~(\ref{#1}) and~(\ref{#2})}
\newcommand{\Figure}[1]{Fig.~\ref{#1}}

\newcommand{\be} {\mathbf{E}}

\newcommand{\br} {\mathbf{r}}

\newcommand{\maxwell} {\ten{\mathcal{M}}}

\newcommand{\id}{\text{I}}
\newcommand{\ten}[1]{\underline{\underline{#1}}}
\newcommand{\G}[1]{\ten{G}_{\,#1}}

\begin{document}

\title{Ab initio theory of Fano resonances in plasmonic nanostructures and metamaterials}

\author{Benjamin Gallinet, Olivier J. F. Martin}
\affiliation{Nanophotonics and Metrology Laboratory, Swiss Federal Institute of technology (EPFL), CH-1015 Lausanne}
\email{benjamin.gallinet@epfl.ch, olivier.martin@epfl.ch}
\begin{abstract}
An ab initio theory for Fano resonances in plasmonic nanostructures and metamaterials is developed using Feshbach formalism. It reveals the role played by the electromagnetic modes and material losses in the system, and enables the engineering of Fano resonances in arbitrary geometries. A general formula for the asymmetric resonance in a non-conservative system is derived. The influence of the electromagnetic interactions on the resonance line shape is discussed and it is shown that intrinsic losses drive the resonance contrast, while its width is mostly determined by the coupling strength between the non-radiative mode and the continuum. The analytical model is in perfect agreement with numerical simulations.
\end{abstract}


\maketitle
\section{Introduction}
Asymmetric resonances display unique features, compared to their symmetric Lorentzian counterpart, and are currently the subject of considerable research efforts in photonic and plasmonic nanostructures. A theoretical derivation was first proposed by Fano to explain autoionization of atoms and the asymmetric shape of these resonances that now bear his name~\cite{Fan61}. In fact, the interference phenomenon underlying Fano resonances is a general wave phenomenon, appearing in particular as Wood anomalies in gratings~\cite{Sar03}, in extraordinary optical transmission~\cite{Aba07}, dielectric~\cite{Fan02} and metallic~\cite{Chr03} photonic crystals, and more recently in optomechanical systems~\cite{Wei10}, plasmonic nanostructures~\cite{Chr08,Ver09,Fan10,Luk10,Son10}, or as the plasmonic analog of electromagnetically induced transparency (EIT) in metamaterials~\cite{Zha08,Liu09,Liu10}. Fano resonances exhibit a very strong sensitivity to changes of the local environment as well as a sharp spectral dispersion. The publications of Luk'yanchuk $\textit{et al.}$~\cite{Luk10} and Miroshnichenko \textit{et al.}~\cite{Mir10} provide extensive reviews on the state of the art of Fano resonances in nanoscale structures and their future application prospects, which include sensing, optical modulation and switching or non-linear devices. Although Fano resonances have been studied in a broad variety of complex plasmonic nanostructures and metamaterials, their analysis relies either on a classical oscillator model~\cite{Alz02,Joe06}, phenomenological models~\cite{Fan02,Chr07}, coupled-mode formalism~\cite{Rua10} or the quantum mechanical approach used by Fano~\cite{Fan61} to fit experimental data and to understand the mechanisms behind the resonance shape. For example, it is not well understood in a realistic plasmonic system composed of several interacting particles how the individual modes and their coupling affect the overall Fano-like resonance of the system. The fact that a discrete radiative mode supported by a plasmonic nanoparticle can act as a continuum in Fano interferences still remains a challenging theoretical task.

\begin{figure}
\includegraphics{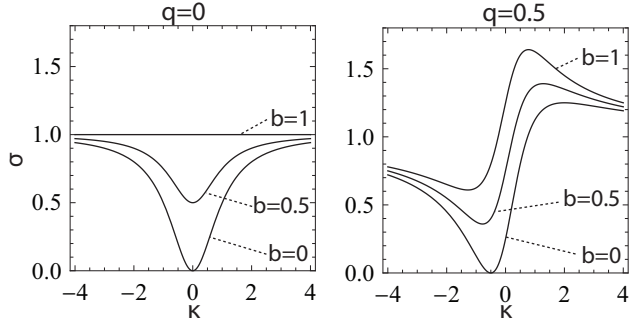}
\caption{Resonance shape function $\sigma/a$ as a function of the reduced frequency $\kappa=(\omega^2-\omega_d^2-\omega_d\Delta)/\Gamma$ for different values of the asymmetry parameter $q$ and the screening parameter $b$ [\Equation{final}]. Intrinsic losses affect the Fano resonances contrast described by the parameter $b$, induce a spectral broadening and a damping of the asymmetry parameter.}
\label{fig1}
\end{figure}

We develop in this work an \textit{ab initio} theory for asymmetric resonances for electromagnetic scattering in general dispersive and lossy media. A general formula for the asymmetric resonance in a non-conservative medium is derived [\Equation{final} and \Figure{fig1}]. This theory reveals the role played by the electromagnetic modes and material losses and enables the engineering of Fano resonances. The influence of electromagnetic coupling onto the resonance line shape is illustrated with the numerical example of a realistic plasmonic nanostructure. It is finally shown that our final result, \Equation{final}, is also valid for the mechanical model of two coupled oscillators.

\section{Electromagnetic theory of Fano resonances}
We consider the electromagnetic scattering on a dielectric or metallic object in a dielectric background. The scatterer supports a non-radiative (dark) mode interacting with a continuum or a radiative (bright) mode. A harmonic time-dependence for the fields ${\bf E}={\bf E}_0e^{-i\omega t}$ is assumed throughout. The relative dielectric permittivity $\epsilon(\br,\omega)$ is generally complex and has frequency dispersion. The frequency dependant electric field $\be$ satisfies the wave equation
\begin{align}
\epsilon^{-1}(\br,\omega)\nabla\times\nabla\times\be(\br,\omega)-\frac{\omega^2}{c^2}\be(\br,\omega)=0\,.
\end{align}
To simplify notation, in the following the dependance of the electric field $\be$ on $\omega$ is kept implicit. Let us assume that the scatterer permittivity is given by a Drude model with plasma frequency $\omega_p$; the scaling law of Maxwell's equations allows then to scale all frequency units by $\omega_p$ and length units by $2\pi c/\omega_p$. We introduce the generally complex and frequency dependant differential operator $\maxwell_\omega$ defined by
\begin{align}
\maxwell_\omega \be(\br)=\frac{c^2}{\epsilon(\br,\omega)}\nabla\times\nabla\times\be(\br)\,.
\end{align}
The wave equation can be written for a vectorial wave function $\ket{\be}$
\begin{align}
\label{baseq}
(\maxwell_\omega-\omega^2\id)\ket{\be}=0\,,
\end{align}
where $\id$ is the identity operator. The inner product is defined by
\begin{align}
\scalar{\be_1}{\be_2}=\int \be_1^*(\br)\cdot\be_2(\br)\text{d}^3\br\,.
\end{align}
Following Feshbach, Bhatia and Temkin, we introduce the orthogonal projection operators $P$ and $Q$ splitting the wave function into a radiative (bright) and a non-radiative (dark) part~\cite{Fes62,Bha84}. Any wave function $\ket{\be}$ can be decomposed as $\ket{\be}=Q\ket{\be}+P\ket{\be}$ where only $P\ket{\be}$ satisfies the radiation condition. Equation~(\ref{baseq}) becomes \begin{align}
(\maxwell_\omega-\omega^2\id)(Q\ket{\be}+P\ket{\be})=0\,,
\end{align}
yielding the two coupled equations:
\begin{align}
\label{coupled1}
(Q\maxwell_\omega Q-\omega^2\id)Q\ket{\be}&=-Q\maxwell_\omega P\ket{\be}\\
\label{coupled2}
(P\maxwell_\omega P-\omega^2\id)P\ket{\be}&=-P\maxwell_\omega Q\ket{\be}\,.
\end{align}
We consider a unique non-radiative mode $\ket{\be_d}$, defined to be eigenfunction of the projector to non-radiative modes $Q\ket{\be_d}=\ket{\be_d}$, and to satisfy $Q\maxwell_{\omega_d}Q\ket{\be_d}=z_d^2\ket{\be_d}$ and $|\scalar{\be_d}{\be_d}|^2=1$. Taking material losses into account, the quantity $z_d=\omega_d+i\gamma_d$ is generally complex. Its real part $\omega_d$ is the mode resonance frequency, and $\gamma_d$ its intrinsic damping. The system is studied in the vicinity of the resonance frequency $\omega_d$. In this frequency range, we assume that the space of eigenfunctions of $Q$ is spanned by $\ket{\be_d}$ only~\cite{Oma65}, so that
\begin{align}
Q=\ket{\be_d}\bra{\be_d}\,.
\end{align}
Equation~(\ref{coupled1}) can be written in $Q$-space as:
\begin{align}
\label{sol}
Q\ket{\be}=\frac{1}{\omega^2-z_d^2}\ket{\be_d}\bra{\be_d}\maxwell_\omega P\ket{\be}\,.
\end{align}
Inserting \Equation{sol} into \Equation{coupled2}, a wave equation with source is obtained for $P\ket{\be}$:
\begin{align}
\label{peq}
(P\maxwell_\omega P-\omega^2\id)P\ket{\be}=\frac{1}{z_d^2-\omega^2}P\maxwell_\omega\ket{\be_d}\bra{\be_d}\maxwell_\omega P\ket{\be}\,.
\end{align}
The radiative (bright) wavefunction $\ket{P\be_b}$ is defined to satisfy \Equation{peq} if the mode $\ket{\be_d}$ is removed. The interference between $\ket{P\be_b}$ and $\ket{\be_d}$ will result in a Fano-like resonance for the total wavefunction $\ket{\be}$. From its definition, $\ket{P\be_b}$ satisfies the following homogeneous equation:
\begin{align}
\label{homeq}
(P\maxwell_\omega P-\omega^2)\ket{P\be_b}=0\,.
\end{align}
Equation~(\ref{peq}) can be solved using the dyadic Green's function $\G{b}$ of \Equation{homeq}:
\begin{align}
\label{solu}
P\ket{\tilde \be}=P\ket{\be_b}+\frac{\bra{\be_d}\maxwell_\omega P\ket{\tilde \be}}{z_d^2-\omega^2}\G{b}P\maxwell_\omega\ket{\be_d}\,.
\end{align}
In \Equation{solu}, the wavefunction $\ket{\tilde \be}$ is related to $\ket{\be}$ but does not have the same asymptotic behavior; an expression for $\ket{\be}$ is derived later. Multiplying \Equation{solu} on the left by $\bra{\be_d}\maxwell_\omega$ and replacing $\bra{\be_d}\maxwell_\omega P\ket{\tilde \be}$ in \Equation{solu} yields:
\begin{align}
\label{delta}
P\ket{\tilde \be}=\ket{P\be_b}+\frac{\bra{\be_d}\maxwell_\omega \ket{P\be_b}}{z_d^2-\omega^2+\omega_d\Delta}\
\G{b}P\maxwell_\omega\ket{\be_d}\,,
\end{align}
where
\begin{align}
\Delta=-\bra{\be_d}\maxwell_\omega P\G{b}P\maxwell_\omega\ket{\be_d}/\omega_d
\end{align}
is the shift in the resonance position $\omega_d$ due to the field overlap between the continuum $\ket{P\be_b}$ and $\ket{\be_d}$. Using \Equations{sol}{solu}, we have a final expression for $\ket{\tilde \be}$:
\begin{align}
\label{finalU}
\ket{\tilde \be}=\ket{P\be_b}+\frac{\bra{\be_d}\maxwell_\omega \ket{P\be_b}}{z_d^2-\omega^2+\omega_d\Delta}(\G{b}P\maxwell_\omega\ket{\be_d}-\ket{\be_d})\,.
\end{align}
The wavefunction $\ket{\be}$ must have the same norm as $P\ket{\be_b}$ in the far-field (i.e. $|P\ket{\be}|^2=|P\ket{\be_b}|^2$). The asymptotic behavior of $\ket{\tilde \be}$ is now compared to $\ket{P\be_b}$, leading to a relation between $\ket{\tilde \be}$ and $\ket{\be}$. Most of plasmonic nanostructures and metamaterials are embedded in a dielectric medium, either homogeneous or on a substrate, allowing us to assume that the permittivity is real and non dispersive in the radiative region. Therefore $P\maxwell_\omega$ is self-adjoint and the set of solutions $\ket{P\be_b}$ of \Equation{homeq} forms an orthogonal basis of modes. The Green's function $\G{b}$ is expanded on this continuum :
\begin{align}
\G{b}=\frac{1}{2\pi}\int \text{d}^3\omega'\frac{\ket{P\be_b(\omega')}\bra{P\be_b(\omega')}}{\omega'^2-\omega^2}\,,
\end{align}
yielding:
\begin{align}
\label{asymp}
P\ket{\tilde \be}=\ket{P\be_b}\left[1-\frac{|\bra{\be_d}\maxwell_\omega \ket{P\be_b}|^2}{2\omega(z_d^2-\omega^2+\omega_d\Delta)}i\right]\
\,.
\end{align}
Let us define the intrinsic damping parameter:
\begin{align}
\Gamma_i=\frac{|\bra{\be_d}\maxwell_\omega \ket{P\be_b}|^2\gamma_d \omega_d}{\omega(\omega_d^2-\omega^2+\omega_d\Delta)^2}\,,
\end{align}
the resonance width:
\begin{align}
\Gamma=\frac{|\bra{\be_d}\maxwell_\omega \ket{P\be_b}|^2}{2\omega(1-\Gamma_i)}\,,
\end{align}
and the reduced frequency:
\begin{align}
\kappa=(\omega^2-\omega_d^2-\omega_d\Delta)/\Gamma\,.
\end{align}
In order for $\ket{\be}$ to have the same normalization in the far-field as $\ket{\be_b}$, it is related to $\ket{\tilde \be}$ by
\begin{align}
\label{eta}
\ket{\be}=\cos\eta\ket{\tilde \be}/(1-\Gamma_i)\,,
 \end{align}
where $\text{cotan}\,\eta=\,\kappa$, considering $\gamma_d\ll \omega_d$ and neglecting any second order contribution. We have asymptotically $P\ket{\be}=P\ket{\be_b}\exp(i\eta)$, meaning that the phase of $\ket{\be}$ shifts rapidly by $\sim \pi$ in a frequency region $\Gamma$ around the resonance position $\omega_d^2+\omega_d\Delta$, resulting in a Fano-like interference. The parameters $\omega_d$ and $\Delta$ determine its spectral position. Its width $\Gamma$ is influenced by the field overlap between the continuum $\ket{P\be_b}$ and the non-radiative mode $\ket{\be_d}$, which is a direct effect of their interference. We consider now a transition operator $T$ from an initial excited state $\ket{g}$ to a final state $\ket{\tilde \be}$ with emission of a photon, which can be interpreted as the response of an optical system to an external excitation, such as the Local Density of States (LDOS)~\cite{Nov06}, the forward scattering cross-section~\cite{Boh83} or the reflectance of a two-dimensional array. From \Equation{finalU}:
\begin{align}
\label{almost}
\frac{\bra{g}T\ket{\tilde \be}}{\bra{g}T\ket{P\be_b}}=1-\frac{|\bra{\be_d}\maxwell_\omega \ket{P\be_b}|^2}{2\omega(z_d^2-\omega^2+\omega_d\Delta)(1-\Gamma_i)}q\,,
\end{align}
with
\begin{align}
\label{q}
q=2\omega(1-\Gamma_i)\frac{\bra{g}T\ket{\be_d}-\bra{g}T\G{b}P\maxwell_\omega\ket{\be_d}}{(\bra{\be_d}\maxwell_\omega \ket{P\be_b})^*\bra{g}T\ket{P\be_b}}\,.
\end{align}
The parameter $q$ is given by the ratio between the optical response of the perturbed non-radiative mode and the continuum. Assuming $q$ is real, one gets from \Equations{eta}{almost} the ratio of the optical response of the total field $\ket{\be}$ to the one of the continuum $\ket{P\be_b}$:
\begin{align}
\label{final}
\sigma=\frac{|\bra{g}T\ket{\be}|^2}{|\bra{g}T\ket{P\be_b}|^2}=a\frac{(\kappa+q)^2+b}{\kappa^2+1}\,,
\end{align}
where $b=\kappa^2\Gamma_i^2q^2(1-\Gamma_i)^{-2}$ and $a=(1-\Gamma_i)^{-2}$. Equation~(\ref{final}) is a generalization of Fano formula~\cite{Fan61,Bha84} to vectorial fields and lossy materials. It describes in particular the resonance strength in plasmonic nanostructures. If the non-radiative mode has no intrinsic losses, one has $\Gamma_i=0$, $b=0$ and the Fano formula is recovered. The parameters $\Gamma$, $\Gamma_i$, $q$, $\Delta$ and $b$ in \Equation{final} are assumed to be constant in the resonance region, but one could consider their lowest energy-dependant corrections over a larger range of frequencies~\cite{Bha84}.

The parameter $b$ does not play a role for the spectral width of the resonance but it influences its contrast (\Figure{fig1}), preventing it to reach zero values. Assuming the parameter $b$ is constant over a frequency region around $\omega_d$, it can be evaluated at $\omega=\omega_d$, leading to:
\begin{align}
\label{b}
b\simeq 4\frac{\gamma_d^2q^2}{\Delta^2}\,.
\end{align}
In the denominator of $b$ appears the product $\bra{\be_d}\maxwell_\omega P\G{b}P\maxwell_\omega\ket{\be_d}(\bra{\be_d}\maxwell_\omega \ket{P\be_b})^*$ accounting for the overlap between the non-radiative mode $\ket{\be_d}$ and the continuum $\ket{P\be_b}$. If the coupling between the two modes is too weak compared to intrinsic losses, the parameter $b$ increases, which has for effect to screen the Fano resonance~(\Figure{fig1}). The screening parameter $b$ is a limitation for the resonance width $\Gamma$, which appears to be the main effect of intrinsic losses in Fano resonances. It is also a critical parameter for achieving electromagnetically induced transparency since it determines the maximal amount of light that can be transmitted~\cite{Zha08,Liu09}. Intrinsic losses included in $\Gamma_i$ are also responsible for a damping of the resonance width $\Gamma$ and the asymmetry parameter $q$. The parameters $q$ and $b$ together describe the resonance shape, and depend on what the transition element $T$ is chosen to represent: for instance the local density of states (LDOS)~\cite{Nov06}, the forward scattering cross-section~\cite{Boh83} or the reflectance of a two-dimensional array.

\section{Numerical validation}
\label{numerics}
\begin{figure}
\includegraphics{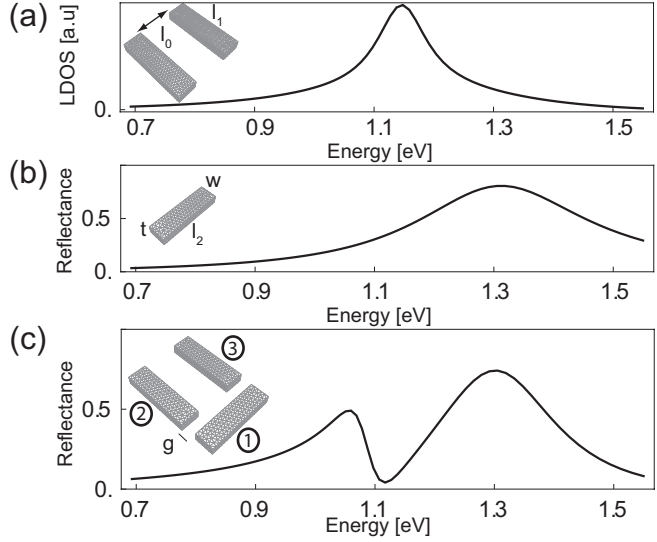}
\caption{Structural decomposition of a Fano resonance in a dolmen-type plasmonic structure in air: beam 1 supports a radiative dipolar mode, whereas beams 2 and 3 support a non-radiative quadrupolar mode. (a) Local density of states (LDOS) of a dipole emitter placed 50nm from the end of one beam of the double beam structure, (b) reflectance of a single beam, (c) reflectance of the composite system. Insets: meshed structures and geometry parameters; $l_0=$160 nm, $l_1=l_2=$300 nm, $w=$80nm and $t=$40nm and $g=30$nm. For the calculation of reflectance spectra, the structures are arranged on a two-dimensional array with period 500nm to avoid neighbor-to-neighbor interactions. A Drude model with plasma frequency
$\omega_p=1.367\times 10^{16}$s$^{-1}$
and damping $\omega_\gamma=0.009\omega_p$ has been chosen for the scatterers'material.}
\label{fig2}
\end{figure}

We now verify numerically the validity of \Equation{final} with the example of a dolmen-type plasmonic structure~\cite{Ver09,Liu09,Liu10}, made of three metallic beams arranged as in \Figure{fig2}. The method used to model electromagnetic scattering on plasmonic structures is based on surface integrals~\cite{Gal10,Gal10a,Ker09}. In the quasistatic approximation, a dark mode cannot be excited by a planewave, but by a rapidly varying field, such as a dipole placed in the near-field of the structure~\cite{Zha10}. The local density of states (LDOS), defined as the imaginary part of the Green's tensor at the position of the dipole, provides all the spectral and scattering information for the structure~\cite{Mar99}. It is computed in \Figure{fig2} at one end of the double beam structure, revealing the existence of a non-radiative quadrupole mode at an energy of $\hbar\omega_d=1.14$eV. The dipolar mode of the perpendicular beam has a larger spectral width due to radiative losses and ensures the coupling of the non-radiative mode to the radiative continuum. The reflectance spectrum of the dolmens array is calculated and divided by the reflectance of an array of single beam in order to obtain the shape function $\sigma$ shown in \Figure{fig1}.
\begin{figure}
\includegraphics{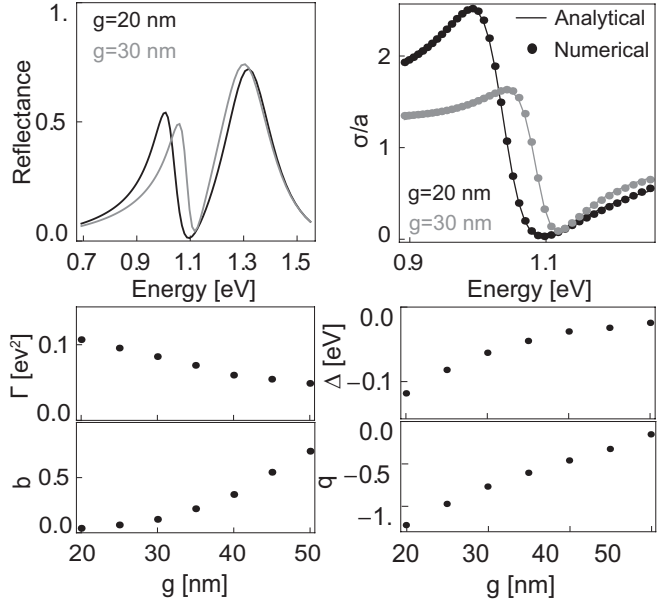}
\caption{Influence of the gap width $g$ on the different parameters of the resonance for the dolmen-type structure in \Figure{fig2}: simulation and fit to \Equation{final}. Each point in the bottom panels corresponds to the fitting parameters extracted from a full calculation for a specific gap distance $g$.}
\label{fig3}
\end{figure}
In \Figure{fig3}, the distance $g$ between beam 1 and the pair of beams 2 and 3 is varied, which changes the strength of the overlap and the coupling between the non-radiative mode and the continuum. For such a realistic plasmonic system for which no analytical solution exist, its optical response does satisfy \Equation{final}. This equation is in perfect agreement with numerical calculations, and allows us to study and engineer its different parameters. As expected from \Equation{delta}, the resonance shift $\Delta$ decreases with decreasing coupling , along with its width $\Gamma$, and the absolute value of the asymmetry parameter $q$ decreases. The coupling strength is therefore a critical parameter for engineering the width of a Fano resonance.
\begin{figure}
\includegraphics{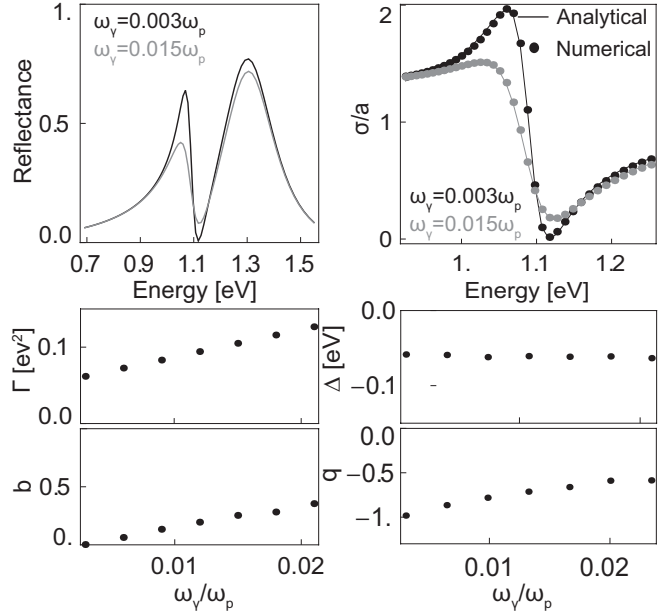}
\caption{Influence of the Drude damping $\omega_\gamma$ on the different parameters of the resonance for the dolmen-type structure in \Figure{fig2}: simulation and fit to \Equation{final}. Each point in the bottom panels corresponds to the fitting parameters extracted from a full calculation for a specific Drude damping $\omega_\gamma$.}
\label{fig4}
\end{figure}
In \Figure{fig4}, the Drude damping $\omega_\gamma$ of the beams 2 and 3 is varied, affecting the imaginary part of the non-radiative mode's eigenvalue $\gamma_d$. Of particular importance is the invariance of the resonance shift $\Delta$, in agreement with the theory which has shown that it is independent on $\gamma_d$. The resonance width $\Gamma$ increases with increasing losses, along with the $b$ parameter; and the $q$ parameter is damped. Overall, intrinsic losses affect the resonance contrast. In order to engineer a sharp and contrasted Fano resonance, an optimal value for both $\Gamma$ and $b$ has to be found.

\section{Mechanical analog: two coupled oscillators model}
In this Section, we show that the final result \Equation{final}, that we have derived from electromagnetic considerations, can also be obtained for the case of two harmonic oscillators with losses.
\begin{figure}
\includegraphics{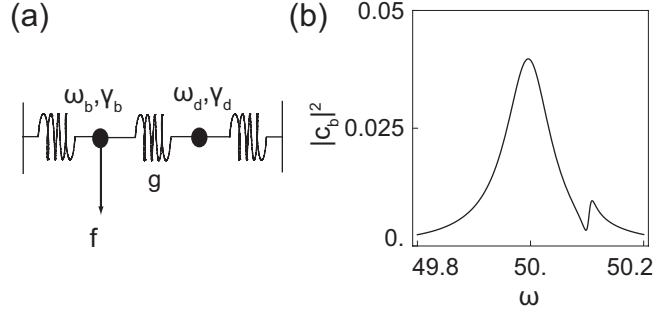}
\caption{(a) Mechanical model of Fano resonances: two coupled oscillators of resonance frequencies $\omega_b$ and $\omega_d$, and damping $\gamma_b$ and $\gamma_d$. One of them is forced by an external excitation of amplitude $f$. The coupling constant is $g$. (b) Amplitude of the forced oscillator as a function of the excitation frequency for $\omega_b=50.$, $\omega_d=50.1$, $g=2.0$, $\gamma_b=0.1$ and $\gamma_d=0.01$.}
\label{fig5}
\end{figure}
Consider the two harmonic oscillators system in \Figure{fig5} which model classically the Fano resonance~\cite{Alz02,Joe06}. The first oscillator modeling the bright (radiative) mode has a resonance frequency $\omega_b$ and high damping $\gamma_b$ representing radiative losses. The second oscillator modeling the non-radiative (dark) mode has a resonance frequency $\omega_d$ and low damping $\gamma_d$. The two oscillators have a coupling $g$. The highly damped oscillator is driven by a harmonic external force with amplitude $fe^{i\omega t}$, representing the coupling to an external field. The equations of motion can be written as:
\begin{align}
\ddot x_b+\gamma_b\dot x_b+\omega_b^2x_b+gx_d&=fe^{i\omega t}\\
\ddot x_d+\gamma_d\dot x_d+\omega_d^2x_d+gx_b&=0\,.
\end{align}
The displacement $x_b$ and $x_d$ of the oscillators is harmonic, therefore $x_b=c_be^{i\omega t}$ and $x_d=c_de^{i\omega t}$. The amplitude of the highly damped oscillator is given by:
\begin{align}
\label{amp}
c_b=\frac{\omega_d^2+i\gamma_d\omega-\omega^2}{(\omega_b^2+i\gamma_b\omega-\omega^2)(\omega_d^2+i\gamma_d\omega-\omega^2)-g^2}f\,.
\end{align}
We now develop the denominator of \Equation{amp} around $\omega_d$, the frequency around which the two oscillators interfere, giving rise to an asymmetric resonance. Considering that $\gamma_d\ll\gamma_b\ll\omega_b,\omega_d$, the quantity $\omega_b^2+i\gamma_b\omega-\omega^2$ is slowly varying in a small frequency interval around $\omega_d$ and can therefore be considered as constant [\Figure{fig5}(b)]. It is evaluated as $C\equiv\omega_b^2+i\gamma_b\omega_d-\omega_d^2$. Around $\omega_d$, Equation~(\ref{amp}) becomes:
\begin{align}
\label{amp2}
c_b\simeq\frac{\omega_d^2+i\gamma_d\omega-\omega^2}{C(\omega_d^2+i\gamma_d\omega-\omega^2)-g^2}f\,.
\end{align}
Neglecting second order contributions from $\gamma_d$, one obtains from \Equation{amp2}:
\begin{align}
\label{finalC}
\frac{|c_b|^2}{|A|^2}=\frac{(\kappa+q)^2+b}{\kappa^2+1}\,,
\end{align}
where $|A|^2=|f|^2/|C|^2$ is the amplitude of the forced oscillator if there is no coupling; $\kappa=(\omega^2-\omega_d^2-\omega_d\Delta)/\Gamma$ is the reduced frequency; $\Delta=[(\omega_d^2-\omega_b^2)g^2]/(|C|^2\omega_d)$ is the resonance shift; $\Gamma=\gamma_b\omega_dg^2/|C|^2$ its width; $q=(\omega_d^2-\omega_b^2)/(\gamma_b\omega_d)$ the asymmetry parameter and $b=\gamma_d^2|C|^4/(\gamma_b^2g^4)$ the screening parameter. The latter also satisfies $b=\gamma_d^2q^2/\Delta^2$ which is a similar relation between the resonance parameters as in \Equation{b}. Hence, Equation~(\ref{finalC}) is the mechanical equivalent to \Equation{final}.

The mechanical and the electromagnetic models both feature the same four independent parameters $\Delta$, $\Gamma$, $q$ and $b$ describing the asymmetric resonance. The parameter $b$ appears if the intrinsic damping $\gamma_d\neq 0$ and increases if $\gamma_d$ increases, having for effect to screen the resonance (\Figure{fig1}), i.e. decrease its contrast. It should also be noted that an increase of the coupling $g$ leads to an increase of the resonance width $\Gamma$, an increase of the absolute value of the shift $\Delta$ and a decrease of the screening parameter $b$, which is in agreement with the electromagnetic theory. The parameters $\Delta$ and $q$ are function of the relative spectral positions of the two oscillators' resonances $\omega_d^2-\omega_b^2$ and vanish if $\omega_b=\omega_d$.

Finally, let us mention that this section shows that \Equation{final} is not only valid in electromagnetism, but also for any physical system that can be modeled by two classical damped coupled oscillators. It is therefore considered as a general formula for the asymmetric resonance in a non-conservative system. Unlike the electromagnetic theory, Equation~(\ref{finalC}) can not provide an expression of the shape parameters $q$ and $b$ including the detection method (for instance back-scattering or forward cross section). This observation appears as a limitation of the mechanical model, compared to the electromagnetic theory.

\section{Summary}
We have derived an \textit{ab initio} theory for Fano resonances in plasmonic nanostructures and metamaterials using Feshbach formalism. The influence of the electromagnetic interactions on the resonance line shapes has been discussed. The novel formula that describes the resonance line shapes introduces a critical screening parameter driven by the intrinsic losses. The resonance width is mostly determined by the coupling strength between the non-radiative mode and the continuum. The formula can be derived from the classical model of two damped coupled oscillators and is in perfect agreement with numerical simulations of complex plasmonic systems. A deep insight into the study of Fano resonances in metallic nanostructures has been enabled.

Funding from CSEM and CCMX-Fanosense as well as stimulating discussion with M. Schnieper and A. Stuck are gratefully acknowledged.


\end{document}